\documentstyle[twocolumn,aps]{revtex}
\begin{document}

\draft

\title{Power law burst and inter-burst interval distributions in the solar wind: Turbulence or dissipative SOC ?}

\author{M. P. Freeman, N. W. Watkins, D. J. Riley \footnote{Now at University of Cambridge, Cambridge, UK.}}

\address{British Antarctic Survey, High Cross, Madingley Road, Cambridge, CB3 
0ET, U.K.}

\date{\today}

\maketitle

\begin{abstract}
We calculate for the first time the probability density functions (PDFs) $P$ of burst energy 
$e$, duration $T$ and inter-burst interval $\tau$ for a known turbulent system in nature. 
Bursts in the earth-sun component of the Poynting flux at 1 AU in the solar wind were measured 
using the MFI and SWE experiments on the NASA WIND spacecraft. We find $P(e)$ and $P(T)$ to 
be power laws, consistent with self-organised criticality (SOC). We find also a power law form for 
$P(\tau)$ that distinguishes this turbulent cascade from the exponential $P(\tau)$ of ideal SOC, 
but not from some other SOC-like sandpile models. We discuss the implications for the relation 
between SOC and turbulence.
\end{abstract}

\pacs{ }

In their seminal papers \cite{BTW1987,BTW1988}, Bak {\it et al.} (BTW) demonstrated 
that a discrete cellular automaton model of an artificial sandpile had a spatial response to 
slow fuelling that was characterised by a scale-free distribution of energy release events or 
``avalanches" (see also \cite{Katz1986}). Scale invariance was shown by a power law probability 
density function (PDF) $P$ of avalanche area $A$, $P(A) = C A^{-\alpha}$. This scale-invariant 
spatial structure led BTW \cite{BTW1987} to propose the sandpile as a toy model of turbulence 
because, in Kolmogorov turbulence \cite{K41}, long-wavelength, injection-range perturbations 
cause a scale-free forward cascade of energy transport until the dissipation scale is reached 
and therefore one might expect the PDFs of burst quantities in turbulent systems 
to be power laws too. These have recently been shown in burst area $A$ for a generic inverse 
cascade model \cite{Gabrielov1999}, in burst energy $e$ and duration $T$ for both a shell model 
\cite{Boffetta1999} and reduced 2D MHD turbulence simulations \cite{EV1999}, and in peak burst 
power for 1D MHD turbulence \cite{GP1998}.

Boffetta {\it et al.} \cite{Boffetta1999} (hereafter B99) have also shown that the PDF $P$ of inter-burst 
intervals $\tau$ in a shell model of turbulence is a power law too but that this is not so for the BTW 
sandpile in which $P(\tau)$ is exponential. B99 postulated that the power law $P(\tau)$ found for 
solar flares \cite{Pearce1993} was consistent with a shell model of turbulence rather than the BTW sandpile. 
Here we demonstrate for the first time that the predicted avalanche phenomenology (power laws in 
$P(e), P(T)$ and $P(\tau)$) of a shell model of turbulence is observed within a natural system 
- the solar wind - for which there is direct independent evidence of turbulence \cite{Mangeney1991}.

The solar wind is a near-radial supersonic plasma outflow from the solar corona which carries 
with it solar magnetic flux into interplanetary space by virtue of the plasma's very high 
electrical conductivity. In this ideal magnetohydrodynamic (MHD) approximation, the electric 
field $\bf{E}^{'}$ in the rest frame of the moving plasma is given by 
$\bf{E}^{'} = \bf{E} + \bf{v} \times \bf{B} = 0$ from Ohm's law. 
The electromagnetic energy (Poynting) flux $\bf{E} \times \bf{H}$ along the sun-earth line $x$ can be 
approximated by $v \left( B_{y}^{2} + B_{z}^{2} \right) / \mu_{0}$ assuming a radial solar wind. 
This quantity was calculated from ``key parameter" measurements of $\bf B$ and $\mid \bf{v} \mid$ 
from the MFI and SWE experiments, respectively, on the WIND spacecraft \cite{Acuna1995} between 
January 1995 and December 1998 inclusive. The typically 80-100~s averaged measurements of 
$\mid \bf{v} \mid$ were interpolated onto the 46~s time samples of $\bf{B}$.

In the resulting time series, bursts were identified, by the method used in \cite{Consolini1997},
as intervals when the Poynting flux exceeded a given fixed threshold. Thresholds were set at the 
10,20,...90 percentiles of the cumulative probability distribution of the Poynting flux. For each 
threshold, the PDF of the burst energy $e$, burst lifetime $T$, and inter-burst interval $\tau$ was 
calculated, where the burst energy is the sum of the Poynting flux samples over the burst 
lifetime $T$.  The PDFs are shown in Figure~1. The burst energy PDF (top panel) can be 
seen to have a power law region over about 4 orders of magnitude between about $10^{-5}$ 
and $10^{-1}$ J m$^{-2}$. The burst lifetime PDF (second panel) also exhibits a power law region 
and can be fitted by a power law with exponential cut-off similar to that found previously for the 
solar wind $\varepsilon$ function \cite{Freeman2000}. In these respects, the solar wind Poynting 
flux has the avalanche phenomenology common to both the BTW sandpile and turbulence.

The inter-burst interval PDF has been plotted on both a log-log scale (third panel) and a
log-linear scale (bottom panel). It is readily seen that this PDF is a power law rather than 
an exponential. A power law with an exponent of 1.67 is shown by the thick dashed curve in the 
third panel. This power law form distinguishes the solar wind from a system having the 
properties of the BTW sandpile and instead shows it to be consistent with the shell model of 
turbulence used by B99. This is the same result they found for solar flares, for which there 
was not the direct independent evidence of turbulence that there is for the solar wind.

It is possible that the solar wind avalanche phenomenology is simply dominated by the advection of an 
already turbulent fluid from the sun rather than by an energy cascade within the solar wind itself 
(S. C. Chapman, personal communication, 1999). We can expect the solar wind outflow from the sun 
to be strongly influenced by energy dissipation events in the solar corona such as nanoflares 
\cite{Parker1988} because these events can change the thermal pressure gradient that drives the 
solar wind \cite{Parker1958} and/or allow reconfigurations of the solar magnetic field that aid 
or inhibit plasma outflow from the sun. These observations are also topical in magnetospheric 
physics because we have previously shown \cite{Freeman2000} a similarity between the avalanche 
phenomenology present in geomagnetic perturbations \cite{Consolini1997,Freeman2000} (which 
measure dissipative currents in the Earth's ionosphere) and that in the energy delivered by the 
solar wind to the Earth's magnetic and plasma environment. Independently, an analysis of the $R/S$ 
Hurst exponents of  solar wind variables and magnetospheric indices \cite{PN1999} has drawn 
similar conclusions.

So what does the observation of avalanche phenomenology in a natural system tell us about its 
physics? BTW postulated \cite{BTW1987,BTW1988} that the appearance of ``avalanche phenomenology" 
(power law burst PDFs) in Nature was due to an underlying fixed point in the 
dynamics (``criticality") which was attractive (``self-organised") - Self-Organised Criticality 
(SOC). Renormalisation group studies \cite{Pietronero1994} have demonstrated that the 
Abelian BTW model indeed exhibits an attractive fixed point. However, although BTW argued that SOC 
implied avalanche phenomenology, the converse is not true; and in particular the observation of 
avalanche phenomenology in natural systems \cite{Jensen1998} does not by itself prove that such 
systems are SOC.

There are many examples of systems that are either not self-organised or not critical, or both, that 
nevertheless present avalanche phenomenology. Avalanche phenomenology has been seen in the forest fire 
model \cite{Loreto1995} controlled by a repulsive rather than an attractive fixed point; it thus has 
to be tuned to exhibit scaling \cite{Chang1999}. Some other models \cite{Jensen1998} exhibit power-law 
distributions without finite size scaling and so are not {\it bona fide} critical. Avalanche 
phenomenology can also be produced by coherent noise driving \cite{NS1996} or by 
``sweeping of an instability" \cite{Sornette1994}. In addition, the fixed-threshold method of 
estimating burst sizes that was used in \cite{Consolini1997} and the present work may generally 
result in scale-free PDFs if applied to certain types of time series. The action of slicing through 
a fractional Brownian motion (fBm) time series at a fixed level generates a set of crossing times 
known as an isoset, for which the PDF of the time interval between two subsequent crossings has a 
power law form \cite{Addison1997}. Hence the burst duration and inter-burst interval statistics 
drawn from such an fBm time series by the fixed threshold method would also be expected to be 
power laws. 

Clearly it is not sensible to apply the SOC label generally to systems exhibiting avalanche 
{\em phenomenology} \cite{Bak1997}. Instead we should follow B99 in using a restricted definition 
of SOC, implicit in BTW's choice of name, as being the {\em mechanism} of self-organisation to 
a critical state. From this point of view, in order to show the presence of SOC, one has to 
demonstrate those properties of self organization and criticality that are unique to the process 
of SOC rather than simply observing the avalanche phenomena that SOC was designed to account for.

In consequence, the important question remains \cite{EV1999} as to the generality of B99's identification 
of an exponential $P(\tau)$ with the SOC mechanism. Exponential $P(\tau)$ implies that energy release episodes 
are uncorrelated in time because of the standard result that Poisson-distributed random numbers 
have an exponential distribution of waiting times. This will give rise to a $1/f^{2}$ power spectrum 
\cite{Jensen1998} for frequencies higher than those corresponding to the longest correlation time. 
In the BTW model, this is the time for the longest avalanche and is set by the system length. 
Jensen {\it et al.} \cite{JCF1989} found that the BTW system had a $1/f^{2}$ high frequency power 
spectrum in energy flow down the sandpile, rather than the $1/f$ spectrum indicative of long-time 
correlation. 

Whilst exponential $P(\tau)$ certainly holds for the BTW sandpile \cite{Wheatland1998,Boffetta1999}, 
this is not true for some other sandpile models. For example, let us consider the nearest neighbour 
OFC model \cite{CO1992,Jensen1998}. The conservative form of this model has been shown to be critical 
\cite{CP1999} and to evolve to a steady state \cite{Jensen1998}. In this case, $P(\tau)$ is found to 
be exponential \cite{CO1992}. However, there is also a non-conservative form of the nearest neighbour 
OFC model \cite{CO1992,Jensen1998} in which dissipation is introduced. This was recently shown to 
cease to be critical \cite{CP1999} and, in this dissipative case, $P(\tau)$ is found to differ from 
an exponential \cite{CO1992,AM1999}. This supports the identification of exponential $P(\tau)$ with SOC. 

Three classes of sandpile model, all of which modify aspects of BTW SOC, exhibit time correlation 
between bursts - variously reported as a non-exponential $P(\tau)$ in the dissipative OFC model 
\cite{CO1992,AM1999} and as a ``1/f" power spectrum in both running \cite{HK1992} and continuous 
(e.g.\cite{Chau1992}) sandpiles. However, it has yet to be shown that any of these systems are
still SOC in the sense of both posessing an attractive fixed point and showing finite size scaling.

If B99 are correct in identifying time correlation of bursts as a diagnostic for the absence of SOC, 
then there should then be no instance of a model that has an attractive fixed point and finite size 
scaling (self-organized and critical) and which also has time correlated bursts of energy flow 
(specifically a $1/f$ spectrum or nonexponential $P(\tau)$). That is, the time correlation in 
dissipative, running and continuous sandpiles is actually the signature of the breakdown of 
self-organized criticality. The apparent paradox of the observation of scale-free burst PDFs in such 
models is resolved when one recognises that scaling may survive away from the fixed point, and can thus 
co-exist with time correlation \cite{CP1999}. Scaling in both space and time can thus be a robust 
``generic" property of such ``near-SOC" systems even if exact criticality is not. Boffetta et al.'s 
test can then test for the presence of SOC but cannot distinguish any of the modified sandpiles from 
turbulence models, and hence such ``near-SOC" models remain possible descriptions of turbulence.

We are grateful to R. P. Lepping and K. W. Ogilvie for solar wind data from the NASA WIND 
spacecraft. We acknowledge valuable discussions with Sandra Chapman, Iain Coleman,
Tim Horbury, Sean Oughton, Carmen do Prado, Dave Tetreault and Tom Chang. We appreciate the provision 
of preprints by Giuseppe Consolini, Channon Price, Jouni Takalo and Donald Turcotte. NWW acknowledges 
the generous hospitality of MIT.\\

\begin{figure}
\caption{Probability density functions of burst measures for the solar wind 
Poynting flux. 
From top to bottom, the measures are burst energy $e$, duration $T$, and 
inter-burst interval 
$\tau$. The PDFs of all measures have power law regions.}
\end{figure}

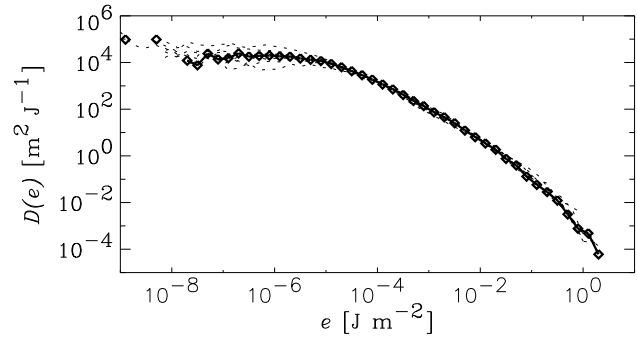
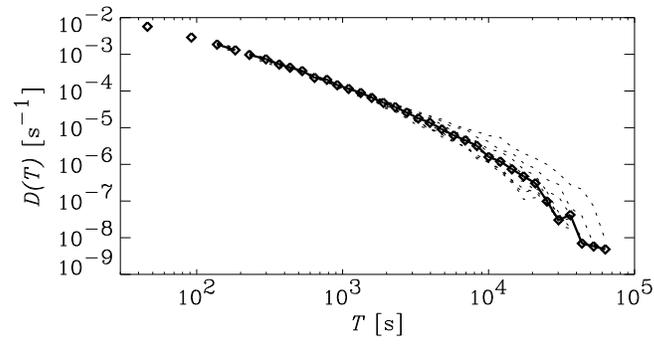
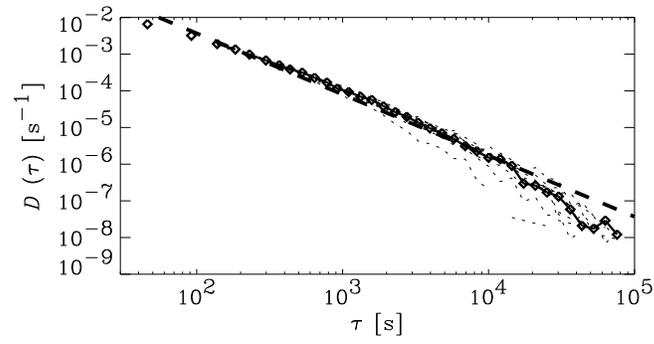
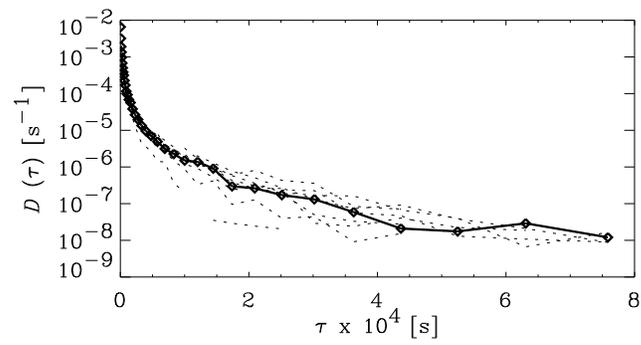
\end{document}